\documentclass[fleqn,usenatbib]{mnras}
\usepackage{newtxtext,newtxmath}
\usepackage[T1]{fontenc}
\usepackage{ae,aecompl}
\usepackage{graphicx}    
\title[Anisotropic SQS]{Anisotropic Strange Stars in the Spotlight: Unveiling Constraints through Observational Data}
\author[Das et al.]{
H. C. Das$^{1}$
\thanks{E-mail: harish.d@iopb.res.in},
Luiz L. Lopes$^{2}$,
\thanks{llopes@cefetmg.br}
\\
$^{1}$ Institute of Physics, Sachivalaya Marg, Bhubaneswar 751005, India\\
$^{2}$ Centro Federal de Educa\c{c}\~ao Tecnol\'ogica de Minas Gerais Campus VIII; CEP 37.022-560, Varginha - MG - Brasil}
\begin{document}
\maketitle
\date{\today}
\begin{abstract}
Motivated by the recent suggestions that very massive pulsar (PSR J0952-0607) and very light compact object (HESS J1731-347) exist, in this article, we revisit the possibility of such objects being strange stars instead of the standard hadronic neutron stars. We study the possible presence of local anisotropy and how it affects the macroscopic properties of strange stars and compare our results with the recent constraints presented in the literature. We found that the presence of anisotropy increases the maximum mass, the radius of the canonical star, and its tidal deformability for positive values of $\lambda_{\rm BL}$ and the opposite for negative values. We also show that although we cannot rule out the possibility of very compact objects being standard hadronic neutron stars, strange stars easily fulfill most of the observational constraints.
\end{abstract}
\section{Introduction}
The last decade saw a great improvement in the physics of the neutron stars. The discovery of the 2.01 $\pm$ 0.04 $M_\odot$ PSR  J0348+0432 pulsar in 2013~\citep{Antoniadis_2013} puts strong constraints in the equation of the state (EOS). However, it was only in the last five years that things started to become very interesting. Results coming from the NASA interior composition explorer (NICER) X-ray telescope and the LIGO/VIRGO gravitational wave observatories constrain not only the masses but also the radii and the tidal deformability of the stars. For instance, the PSR J0740+6620 with a mass of 2.14 $M_\odot$ was pointed in Ref.~\citep{Cromartie_2020}. NICER results refined this data, and today the PSR J0740+6620 with a mass of 2.08 $\pm$ 0.07 $M_\odot$ and a radius of 12.35 $\pm$ 0.35 km ~\citep{Miller_2021} acts as a strong constraint that any valid EOS must fulfill. The canonical star, $M =1.4 M_\odot$, also received great attention in the last years. Two NICER results point that the radius of the canonical stars must be in the range $11.52 <R_{1.4}< 13.85$ km~\citep{Riley_2019}, and $11.96 <R_{1.4}< 14.26$ km~\citep{Miller_2019}. Ultimately, these results were revised to $11.80 <R_{1.4}< 13.10$ km ~\citep{Miller_2021}. Likewise, the LIGO/VIRGO observatories constrain the dimensionless tidal parameter of the canonical star to $\Lambda_{1.4}<800$ in the GW170817 event~\citep{Abbott_2017}. This result was also refined in Ref.~\citep{Abbott_2018}, to  $70<\Lambda_{1.4}<580$.

Beyond these well-established constraints, we can find in the recent literature some extreme and more speculative results. For instance, the black widow pulsars PSR J0952-0607 have been detected in the Milkyway Galaxy, with a measured mass of  $M = 2.35\pm0.17 M_\odot$ \citep{Romani_2022}. If such a high mass is confirmed, the PSR J0952-0607 will stand as the heaviest neutron star ever discovered. An even more extreme case is the mass-gap object in the GW190814 event~\citep{RAbbott_2020}. With an established mass in the range between 2.50 and 2.67 $M_\odot$, the question here is about its true nature (see Ref.~\citep{Lopes_ApJ} for additional discussion). Is it the heaviest neutron star ever or the lightest black hole? The possible nature of the mass-gap object as a degenerate star will not only strongly constrain the mass but also the dimensionless tidal parameter for the canonical star to the range of $458 <\Lambda_{1.4}< 889$. Still on the canonical star, using multimessenger observations, Ref.~\citep{Capano_2020} was able to constrain the radius of the canonical star to the range of only $10.4<R_{1.4}<11.9$ km. An extremely tight bound with a very narrow overlap with the revised NICER one. Finally, in the realm of very light objects, the proper existence of the so-called HESS J1731-347 supernova remnant present a puzzle once it has a unique mass of $M=0.77_{-0.17}^{+0.20}~M_\odot$ and a radius $R = 10.4 _{-0.78}^{+0.86}$ km~\citep{Doroshenko_2022}.

The tight bond for the radius of the canonical star presented by Capano {\it et al.} ~\citep{Capano_2020}) and the existence of the HESS J1731-347 object~\citep{Doroshenko_2022} strongly suggest that at least some of the observed pulsars are strange stars (sometimes called strange quark stars (SQS)) instead of the standard hadronic neutron stars. In this work, we revisit the theory of strange stars, considering not only the isotropic case but also investigating the effects of local anisotropy. The anisotropy in pressure arises due to several exotic phenomena, such as a very high magnetic field \citep{Yazadjiev_2012, Cardall_2001, Ioka_2004, Ciolfi_2010, Ciolfi_2013, Frieben_2012, Pili_2014, Bucciantini_2015}, superfluidity \citep{Kippenhahn_1990, NKGb_1997, Heselberg_2000}, kaon condensations \citep{Sawyer_1972}, phase transitions \citep{Carter_1998}, etc. (see Ref.~\citep{Herrera_1997} for a review ). The first anisotropy model was proposed by the Bower-Liang (BL) in 1974 \citep{Bowers_1974}. There are other models, such as the Horvat model \citep{Horvat_2010} and the Conseza model \citep{Cosenza_1981}, which incorporate anisotropy inside the star, and these models have their own physical significance. For example, the BL-model is based on the assumption that the anisotropy must vanish at the origin, and it should depend non-linearly on the radial pressure \citep{Bowers_1974}; however, the Horvat and Conseza model have their different assumptions. Several studies have already explored to study the anisotropic star and its different properties by varying the degree of anisotropy. For example, Silva {\it et al.} took the limit of the anisotropic parameter $-2\leq \lambda_{\rm BL}\leq +2$ using the BL- model ~\citep{Silva_2015}, and using Horvat model, Doneva {\it et al.} put the limit on $\lambda_{\rm H}$ is from -2 to +2 \citep{Doneva_2012}. One can't arbitrarily choose the degree of anisotropy. From the different observational data, one can put constraints on its degree.

Several studies have already explored the different properties of the anisotropic strange star. For example, the effect of anisotropy on radial oscillation was first studied in Ref. \citep{Arbanil_2016}. The anisotropic charge SQS has been explored in the Tolman–Kuchowicz space-time geometry, where they use the MIT bag model \citep{Maurya_2019}. They got an interesting result that the pressure anisotropy initially dominates the Coulomb repulsive forces, and the repulsive forces dominate the anisotropic when the radius increases. This might require additional forces to equilibrium against the gravitational collapse.  Deb {\it et al.} \citep{Deb_2021} have explored the effect of magnetic field on the anisotropic SQS. They observed that the transverse component of the magnetic field increases the mass and size of the anisotropic star and vice-versa for the radial component of the field. In this study, we explore the properties of anisotropic SQS in light of different observational data within the BL model~\citep{Bowers_1974}. Using different observational data, we want to put a limit on the degree of anisotropicity inside the SQS.  
\section{Formalism}
The theory of strange stars is based on the so-called Bodmer-Witten conjecture~\citep{Bodmer_1971, Witten_1984}, which claims that the ordinary matter we know, composed of protons and neutrons may be only meta-stable, while the true ground state of strongly interacting matter would therefore consist of strange matter (SM), which in turn is composed of deconfined up, down and strange quarks. For the SM hypothesis to be true, the energy per baryon of the deconfined phase (for $p = 0$ and $T = 0$) is lower than the nonstrange infinite baryonic matter, i.e., ($E_{uds}/A < 930$ MeV), while at the same time, the nonstrange matter still needs to have an energy per baryon higher than the one of nonstrange infinite baryonic matter ($E_{ud}/A > 930$ MeV); otherwise, protons and neutrons would decay into $u$ and $d$ quarks.

To model the quark matter, we use a vector-enhanced MIT bag model~\citep{Klahn2015}. Explicitly, we use the thermodynamic consistent vector MIT bag model as introduced in Ref.~\citep{Lopes_2021} whose Lagrangian density reads
\begin{eqnarray}
\mathcal{L}_{\rm vMIT} &=& \bigg\{ \bar{\psi}_q\big[\gamma^\mu(i\partial_\mu - g_{qV} V_\mu) - m_q\big]\psi_q  \nonumber \\
&&
- B + \frac{1}{2}m_V^2V^\mu V_\mu  \bigg\}\Theta(\bar{\psi}_q\psi_q) ,
\label{vMIT}
\end{eqnarray}
where $m_q$ is the mass of the quark $q$ of ﬂavor $u$, $d$ or $s$, $\psi_q$ is the Dirac quark ﬁeld, $B$ is the constant vacuum pressure, and $\Theta(\bar{\psi}_q\psi_q)$ is the Heaviside step function to assure that the quarks exist only conﬁned to the bag.

The parameters utilized in this work are: $m_u = m_d = 4$ MeV, $m_s = 95$ MeV, $B^{1/4} = 145$ MeV, and $G_V = (g_v/m_v)^2 = 0.3$ fm$^2$. Moreover, the obtained EOS is charge neutral and chemical stable. A detailed discussion about the vector MIT bag model, its formalism, and the Bodmer-Witten conjecture can be found in Refs.~\citep{Lopes_2021, Lopes_2021b} and the references therein.

Now, to account for the possible presence of local anisotropy, we define the energy-momentum tensor as ~\citep{Doneva_2012, Silva_2015, Estevez_2018}:
\begin{eqnarray}
    T_{\mu\nu} = (\rho+p_t)u_\mu u_\nu + (p_r-p_t) k_\mu k_\nu + p_t g_{\mu\nu},
    \label{eq:tmunu_aniso}
\end{eqnarray}
where $\rho$, $p_r$, and $p_t$ are the mass density, radial pressure, and tangential pressure, respectively. $k_\mu$ is the unit radial vector ($k^\mu k_\mu = 1$) with $u^\mu k_\mu = 0$. The Schwarzschild metric for this type of star having the spherically symmetric and static configuration is defined as 
\begin{eqnarray}
    ds^2= e^{\nu}dt^2-e^{\lambda}dr^2-r^2d\theta^2-r^2 \sin^2\theta  d\varphi^2\,, \label{Scz}
\end{eqnarray}
where $r$, $\theta$, and $\phi$ are the Schwarzschild coordinates. 

\begin{figure}
    \centering
    \includegraphics[width=0.45\textwidth]{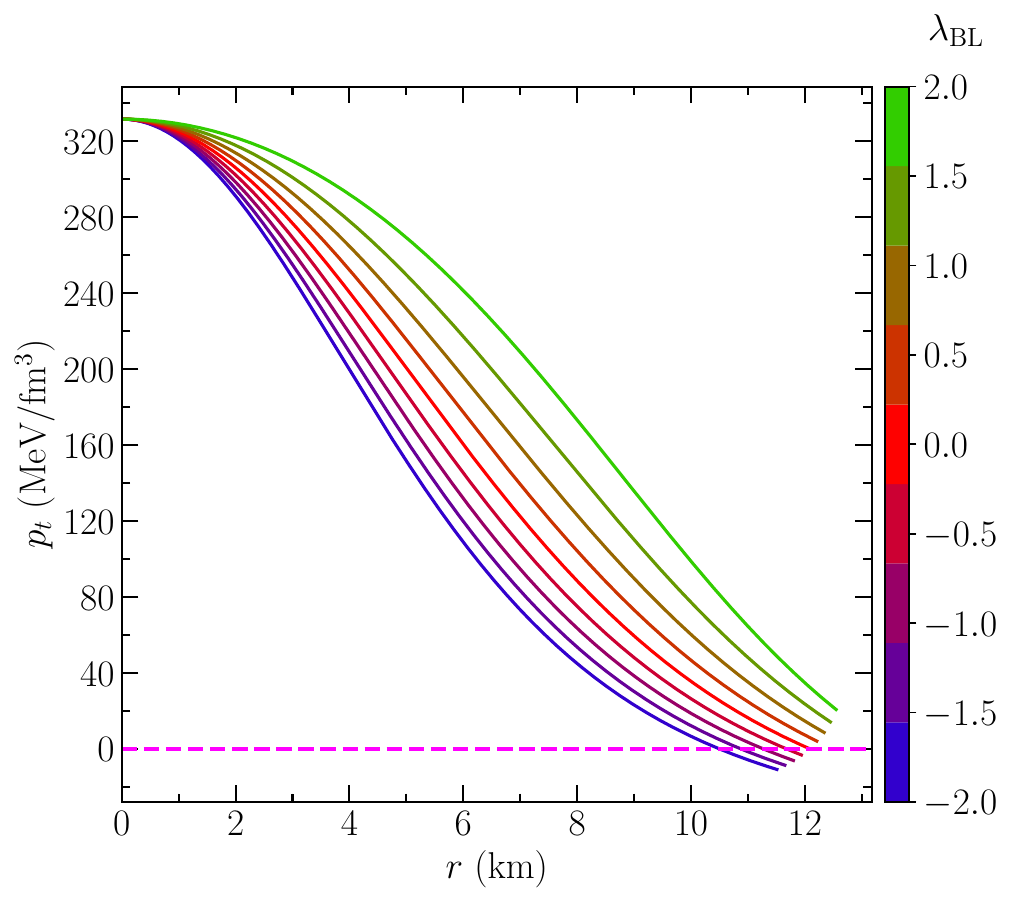} \\
    \includegraphics[width=0.45\textwidth]{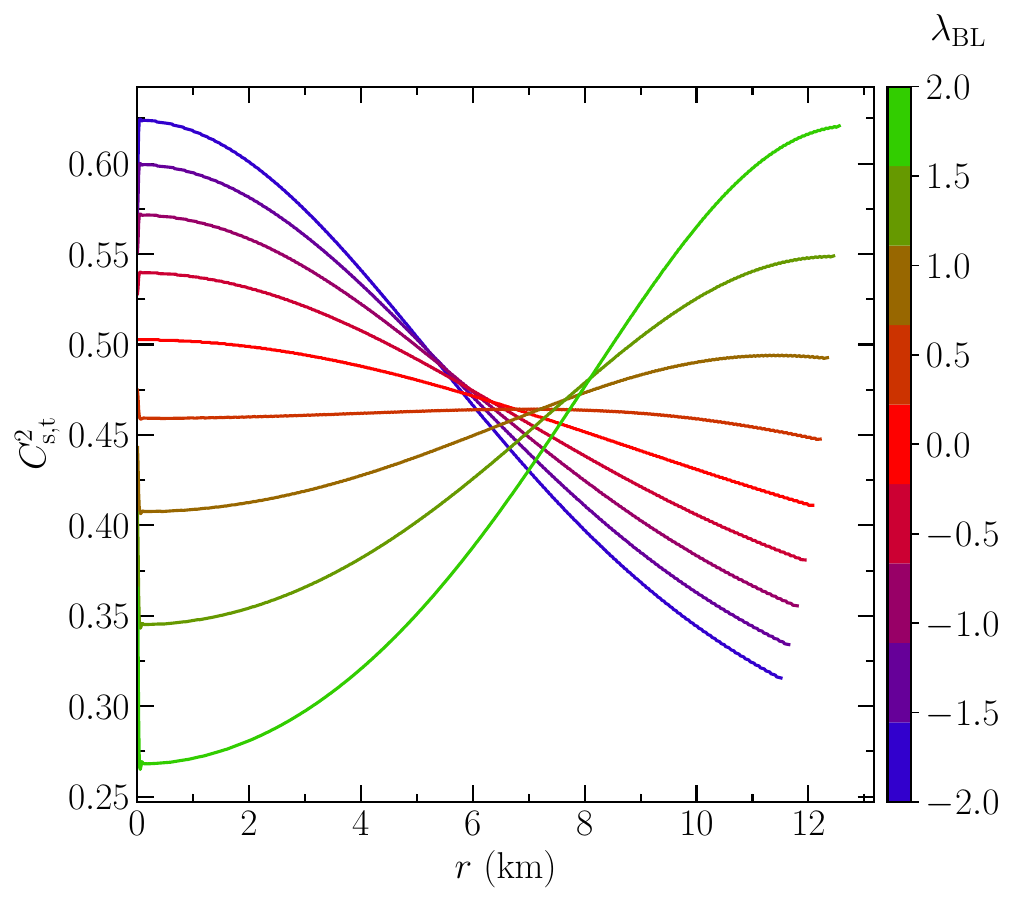}
    \caption{{\it Top:} The tangential pressure as a function of the radius at different values of $\lambda_{\rm BL}$ for the maximum mass of the SQS. {\it Bottom:} The radial variation of the square of the sound speed.}
    \label{fig:press_sigma_css}
\end{figure}

The structure equations for the anisotropic star can be obtained by solving Einstein's equations as ~\citep{Doneva_2012}
\begin{eqnarray}
    \frac{dp_r}{dr}=-\frac{\left( \rho + p_r \right)\left(m + 4\pi r^3 p_r \right)}{r\left(r -2m\right)} +\frac{2\sigma}{r} \,,
    \label{tov1:eps}
\end{eqnarray}
\begin{eqnarray}
    \frac{dm}{dr}=4\pi r^{2}\rho\,,
    \label{tov2:eps}
\end{eqnarray}
where $\sigma=p_t-p_r$ is the anisotropy parameter, `$m$' is the mass enclosed within the radius $r$. The radial pressure is then obtained from a pre-determined EOS.  In this work, we use the vector MIT bag model for SQS obtained from Eq.~\ref{vMIT}. On the other hand, for the case of the transverse pressure, we use the BL model in the following \citep{Bowers_1974}:
\begin{eqnarray}
    \label{Anisotropy_eos}
    p_t = p_r + \frac{\lambda_{\rm BL}}{3} \frac{(\rho+3p_r)(\rho + p_r)r^2}{1-2m/r} \,,
\end{eqnarray}
where the factor $\lambda_{\rm BL}$ measures the degree of anisotropy in the fluid. There are some boundary conditions required to solve the above Eqs. (\ref{tov1:eps}-\ref{Anisotropy_eos}) as done in Refs. \citep{Biswas_2019, Das_ILC_2022}. Also, different fluid conditions must be satisfied for the anisotropic stars such as (i) $p_r, p_t$, and $\rho > 0$, (ii) $0<c_{\rm s, t}^2<1$, etc. Other conditions are mentioned in Refs. \citep{Das_ILC_2022}. However, we also check for some of them here, as shown in Fig. \ref{fig:press_sigma_css} for the maximum mass of the SQS. It is also worth noticing from Eqs.~\ref{eq:tmunu_aniso}--\ref{Anisotropy_eos}, that in the BL model, the presence of local anisotropy does not break the spherical symmetry of the star (see also ~\citet{DelgadoeDelgado2018} to additional discussion).
\section{Results and Discussion}
It is well known that the energy density of the SQS is not zero at the surface, where the pressure vanishes. Therefore, the SQS is also called a self-bound star.  However, the tangential pressure does not vanish for the anisotropic case and even becomes negative for negative values of $\lambda_{\rm BL}$. Moreover, in the case of the standard hadronic neutron stars, both energy density and pressure vanish. Furthermore, as displayed in Fig.~\ref{fig:press_sigma_css}, the tangential pressure decays quicker for negative values of $\lambda_{\rm BL}$ and slowly for the positive ones.  One can see more details in Fig. 1 of Ref. \citep{Das_ILC_2022}.

The square of the speed of sound corresponding to tangential pressure for the anisotropic SQS is also shown in Fig.~\ref{fig:press_sigma_css}, where a peculiar behavior can be observed. The  $C^2_{s,t}$ increases with the radius for positive values of  $\lambda_{\rm BL}$ due to the contribution of the radius and the mass as can be seen in Eq.~\ref{Anisotropy_eos}.
For the isotropic case ($\lambda_{\rm BL}$ = 0), we have a discrete decrease in the speed of sound, while for negative values of $\lambda_{\rm BL}$, we have a quick decrease of the speed of sound. Nevertheless, for all values of $\lambda_{\rm BL}$, we have a  physical and positive solution for the $C^2_{s,t}$. This is an interesting fact about the anisotropic strange star; once for the standard neutron stars, the square of the speed of sound goes to negative for the negative values of $\lambda_{\rm BL}$  as mentioned in Refs. \citep{Biswas_2019, Das_ILC_2022, Silva_2015}.
\begin{figure}
    \centering
    \includegraphics[width=0.45\textwidth]{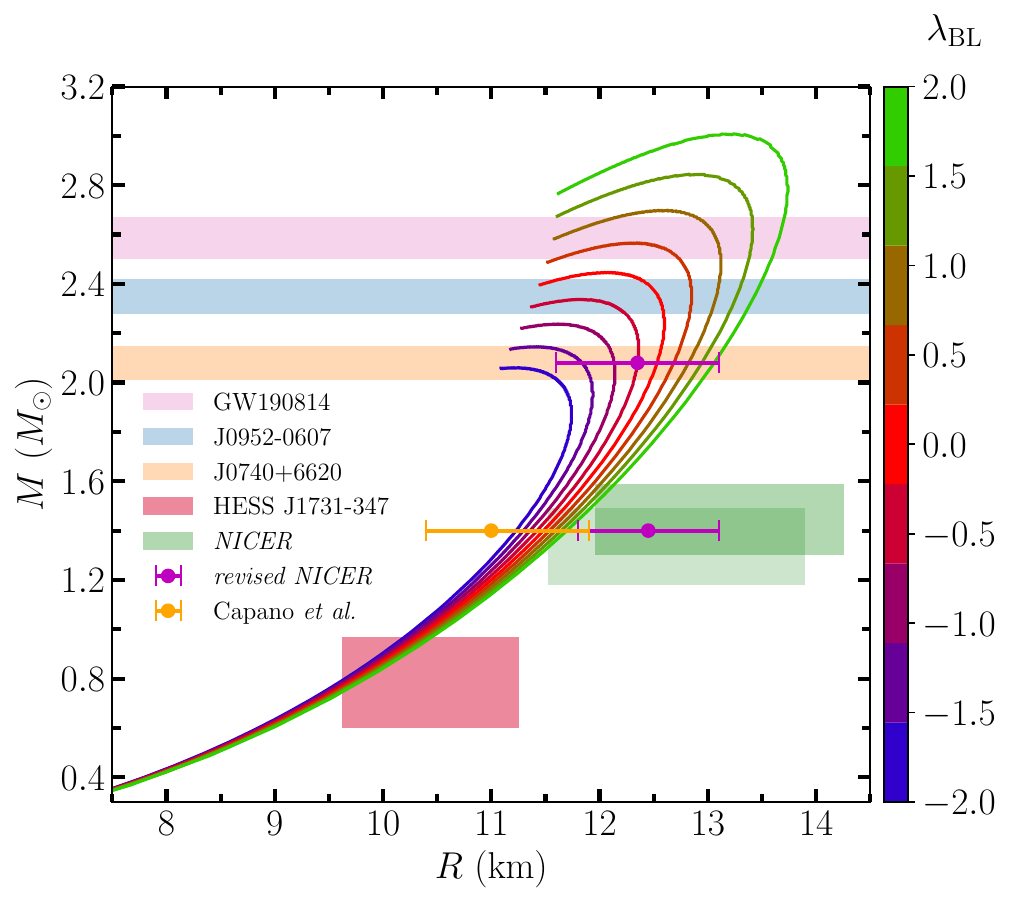} \\
    \includegraphics[width=0.45\textwidth]{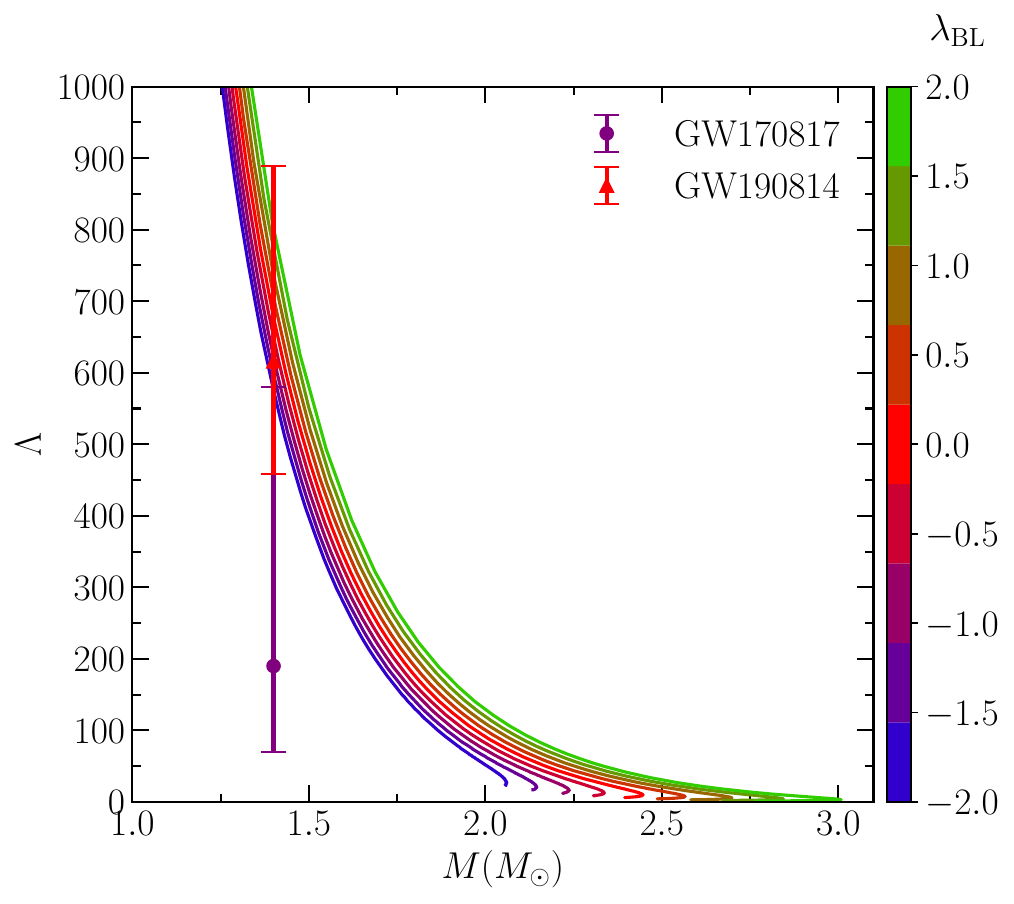}
    \caption{{\it Top:} Mass-radius profiles for anisotropic SQS with $-2.0<\lambda_{\rm BL}<+2.0$ for considered EOS and the constraints discussed in the text. {\it Bottom:} The dimensionless tidal deformability as a function of mass for different $\lambda_{\rm BL}$. The error bars are the observational constraints given by LIGO/Virgo events GW170817 and GW190814.}
    \label{fig:mr_Lambda}
\end{figure}

We solve the structural equations for anisotropic stars and display the mass-radius diagram for different degrees of anisotropy in Fig.~\ref{fig:mr_Lambda}, altogether with the constraints pointed out in the introduction of the paper. Our focus here is the constraints that are difficult to reconcile with standard neutron stars. Explicitly, the $M = 2.35 \pm 0.17 M_\odot$ PSR J0952-0607 black widow pulsar~\citep{Romani_2022}, the radius determined by Capano {\it et al.} ~\citep{Capano_2020}, and the HESS J1731-347 object~\citep{Doroshenko_2022}. Nevertheless, the constraints coming from  NICER~\citep{Miller_2019,Riley_2019,Miller_2021}, and  from the GW190814 event~\citep{RAbbott_2020} are also included. 

We first notice that the effects of the anisotropy are to increase the maximum mass for positive values of $\lambda_{\rm BL}$, and reduce the maximum mass for negative values of $\lambda_{\rm BL}$. The same is true for the radius of the canonical $1.4 M_\odot$ star. Concerning the constraints, we see that for all values of anisotropy, the mass and radius range of the HESS J1731-347 object~\citep{Doroshenko_2022} can be explained by assuming it is a strange star. The same is true for the radius of the canonical star represented by Capano {\it et al.} ~\citep{Capano_2020}, which is bounded by $R_{1.4} < 11.9$ km.  The $M = 2.35 \pm 0.17 M_\odot$ PSR J0952-0607 black widow pulsar~\citep{Romani_2022}, can also be explained by assuming an anisotropy in the range $-1.0~<\lambda_{\rm BL}~< 2.0$. Even the lower limit of $2.50 M_\odot$ in the GW190814 event can be reached if we use $0.5 ~<\lambda_{\rm BL}~< 2.0$. Indeed, within this range for the  $\lambda_{\rm BL}$, we can simultaneously satisfy the constraints coming from the HESS object, the Capano limit, the PSR J0952-0607 black widow pulsar, and the GW190814 event, altogether with the 2.08$~\pm~0.07~M_\odot$  PSR J0740+6620. Unfortunately, all of the 1.4 $M_\odot$ strange stars present a radius below the inferior limit of 11.8 km presented in the revised NICER analysis~\citep {Miller_2021}, yet the original range is still satisfied for 0.5 $~<\lambda_{\rm BL}~<$ 2.0~\citep{Riley_2019}.

We now discuss how the anisotropy affects the dimensionless tidal parameter $\Lambda$. The tidal deformability of a compact object is a single parameter that quantifies how easily the object is deformed when subjected to an external gravitational field. Larger tidal deformability indicates that the object is easily deformable. On the opposite side, a compact object with a smaller tidal deformability parameter is smaller, more compact, and it is more difficult to deform. It is defined as~\citep{Hinderer_2008,Hinderer_2009,Chatziioannou_2020,Flores_2020}:
\begin{equation}
    \Lambda = \frac{2k_2}{3C^5},
\end{equation}
where $C = M/R$ is the compactness of the star. The parameter $k_2$ is called the Love number and is related to the metric perturbation. A detailed calculation of the parameter $k_2$ in the presence of anisotropy for the BL model utilized in this work can be found in Refs.~\citep{Biswas_2019, Das_ILC_2022}. The results are displayed in Fig.~\ref{fig:mr_Lambda} altogether with the constraints coming from the GW17017~\citep{Abbott_2018} and GW190814~\citep{RAbbott_2020} events.

As in the case of the maximum mass, the effects of the anisotropy are to increase the dimensionless tidal parameter for positive values of $\lambda_{\rm BL}$, and reduce it for negative values of $\lambda_{\rm BL}$. About the constraints, we see that only for  $\lambda_{\rm BL}$ = -2.0, the bound of the revised GW170817 event is satisfied, although virtually all values of $\lambda_{\rm BL}$ satisfy the original bound $\Lambda~<$ 800. Furthermore, the GW190814 data limit is well satisfied for all values of $\lambda_{\rm BL}$.  

We summarize the main results for different values of $\lambda_{\rm BL}$ in Tab.~\ref{tab:table} for different values of $\lambda_{\rm BL}$. 
\begin{table}
\caption{The maximum mass ($M$), and its corresponding radius ($R$), canonical radius ($R_{1.4}$). and the dimensionless tidal deformability ($\Lambda_{1.4}$) are enumerated with different values of $\lambda_{\rm BL}$ for assumed EOS. }
\label{tab:table}
\renewcommand{\tabcolsep}{0.3cm}
\renewcommand{\arraystretch}{1.2}
\scalebox{1.0}{
\begin{tabular}{llllll}
\hline \hline
 $\lambda_{\rm BL}$ &
 \begin{tabular}[c]{@{}l@{}} \hspace{0.2cm}$M$ \\ ($M_\odot$)\end{tabular} &
  \begin{tabular}[c]{@{}l@{}} \hspace{0.2cm}$R$ \\ (km)\end{tabular} &
  \begin{tabular}[c]{@{}l@{}}$R_{1.4}$\\ (km) \end{tabular} &
  $\Lambda_{1.4}$ \\ \hline
 +2.0 &  3.00 & 13.24 & 11.70 & 802 \\ \hline
 +1.5 &  2.84 & 12.83 & 11.65 & 774 \\ \hline
 +1.0 &  2.69 & 12.55 & 11.59 & 737 \\ \hline
 +0.5 &  2.57 & 13.35 & 11.54 & 708 \\ \hline
  0.0 &  2.45 & 12.08 & 11.48 & 678 \\ \hline
 -0.5 &  2.34 & 11.82 & 11.42 & 650 \\ \hline
 -1.0 &  2.24 & 11.62 & 11.36 & 623 \\ \hline
 -1.5 &  2.15 & 11.42 & 11.30 & 598 \\ \hline
 -2.0 &  2.06 & 11.25 & 11.23 & 575 \\ \hline \hline
\end{tabular}}
\end{table}
\section{Conclusion}
In this article, we explored some of the macroscopic properties of the anisotropic SQS. The local anisotropy in the pressure inside the SQS has been included with the help of the BL model. Although there are other models which could incorporate the anisotropy inside the star, the BL model is simple and assumes that the anisotropy is gravitationally induced and varies non-linearly with radial pressure. Using the BL model, we have calculated some of the perfect fluid conditions as shown in Fig. \ref{fig:press_sigma_css}. From this figure, it has been observed that the tangential pressure in the case of an isotropic star is zero at the surface; however, the tangential pressure is not null at the surface with the inclusion of anisotropy. Similarly, we have observed that the energy density is not null at the surface of the SQS. The speed of sound also obeys the causality condition throughout the star with the inclusion of anisotropy. It is a fascinating fact about the SQS in comparison to NS. 

The SQS EOS has been obtained with the vector MIT bag model with a suitable choice of bag constant, masses of the quarks, and coupling constants in order to satisfy the Bodmer-Witten conjecture.  With this obtained EOS, we have calculated the mass-radius profiles for anisotropic SQS by varying the degree of anisotropy inside the star. Some of the important observational data are overlaid, which helps us to put a constraint on the degree of anisotropy. The heaviest pulsar PSR J0740+6620  mass limit has been satisfied even with the lower value of $\lambda_{\rm BL}$. The higher values of anisotropy reach the limits given by GW190814 and the PSR J0952-0607 black widow pulsar. Unfortunately, no curves passed through the revised NICER data for the canonical star. On the other hand, all curves are in agreement with the tight bound pointed by Capano {\it et al.}. Other properties such as $\Lambda$ have also been studied. It has been observed that all curves pass through the GW190814 data, which pointed out that the secondary object might be an anisotropic SQS.  

Ultimately, the existence of the  HESS J1731-347 object puts strong constraints on the theory of neutron stars. Nevertheless, the possibility of the HESS object being a standard neutron star cannot be ruled out. In~\citet{Doroshenko_2022}, the authors show that some EoS derived from Chiral perturbation theory are capable of fulfilling not only HESS J1731-347 but many other compact star constraints. On the other hand, in~\cite{Sagun2023}, the authors show that the HESS object can be a hadronic neutron star with a soft EoS, or a hybrid star with an early deconfinement phase transition. A possible hadronic nature was also explored in~\cite{huang_2023} and \cite{Kubis_2023}. Another possibility one can not exclude is that the HESS object might be a dark matter admixed neutron star as mentioned in Ref. \citep{Routaray_2023}. Here, our theoretical predictions satisfy both mass-radius constraints provided by HESS and simultaneously well-reproduced other observational data. {\it Therefore, we suggest that the HESS object might be an anisotropic SQS}. In future observations, one can limit both degrees of anisotropy and the type of compact object. We hope that other theoretical predictions based on different anisotropy models could also help us to study the nature of HESS objects.
\section*{DATA AVAILABILITY}
This manuscript has no associated data, or the data will not be deposited. Data sharing is not applicable to this article as no data sets were generated or analyzed during the current study.
\bibliography{anisotropic_SQS}
\bibliographystyle{mnras}
\end{document}